\documentclass[prl,twocolumn]{revtex4}
\usepackage{graphicx}
\usepackage{dcolumn}
\usepackage{bm}
\usepackage{amsmath}
\usepackage{amssymb}
\usepackage{mathrsfs}
\usepackage{amsfonts}

\begin{document} 

\title{Controlling discrete and continuous symmetries in 'superradiant' phase transitions}

\author{Alexandre Baksic}
\email{alexandre.baksic@univ-paris-diderot.fr}
\author{Cristiano Ciuti}
\email{cristiano.ciuti@univ-paris-diderot.fr}
\affiliation{Laboratoire Mat\'eriaux et Ph\'enom\`enes Quantiques,
Universit\'e Paris Diderot-Paris 7 and CNRS, \\ B\^atiment Condorcet, 10 rue
Alice Domon et L\'eonie Duquet, 75205 Paris Cedex 13, France}

\begin{abstract}
We explore theoretically the physics of a collection of  two-level systems coupled to a single-mode bosonic field in the non-standard configuration where each (artificial) atom is coupled to both field quadratures of the boson mode.
We determine the rich phase diagram showing 'superradiant' phases with  different symmetries. We demonstrate that it is possible to pass from a discrete, parity-like $\mathbb{Z}_2$ symmetry to a continuous $U(1)$ symmetry even in the ultrastrong coupling regime where the rotating wave approximation for the interaction between field and two-level systems is no longer applicable. By applying this general paradigm, we propose a scheme for the experimental implementation of such continuous $U(1)$ symmetry in circuit QED systems, with the appearance of photonic Goldstone and amplitude modes above a critical point. 
\end{abstract}

\maketitle

The collective 'superradiant' coupling of a large number of two-level systems to a bosonic field has been attracting a remarkable interest since the pioneering paper by Dicke\cite{Dicke} and is now the focus of many recent studies in cavity\cite{Dimer,esslinger,Nagy,Baumann,Bhaseen} and circuit \cite{NatafPRL1,NatafNAT,NatafPRL2,NatafPRA,Baksic} quantum electrodynamics (QED).
In particular, the well-known Dicke model describes the coupling between a collection of two-level systems and a single photon mode. For increasing atom-field
coupling such a model predicts a superradiant phase transition \cite{Lieb,Carmichael,Brandes}, with a doubly degenerate ground state above a critical value of the vacuum Rabi coupling. 
The so-called 'superradiant' phase ground state is characterized by a spontaneous polarization of the two-level systems and a spontaneous coherence of the boson field. The Dicke Hamiltonian has a discrete $\mathbb{Z}_2$-symmetry: there is no continuous $U(1)$ symmetry due to the so-called non-rotating wave terms of the interaction between the two-level systems and the field, which cannot be neglected in the so-called ultrastrong coupling regime \cite{CiutiPRB2005,Anappara,Niemczyk} (a recent work \cite{Yixiang} instead considered a model neglecting such non-rotating wave terms, which are however important for the large couplings required to have 
a superradiant phase transition \cite{Carmichael,Brandes}). Recent judicious generalizations of the Dicke model have been explored to control the corresponding Hamiltonian symmetry\cite{NatafPRA}, which however remains still discrete.

Phase transitions with artificial systems having a continuous $U(1)$ symmetry are attracting a significant interest, for example in Bose-Hubbard systems exploiting ultracold atoms\cite{Bloch}, due to the connections with the exciting physics of the Anderson-Higgs mechanism\cite{Anderson,Higgs}. In this letter, we explore a model describing a collection of two-level systems, each one coupled to both quadratures of a boson mode.
We show that by tuning the two quadrature coupling constants it is possible to control the symmetries of the system, with the possibility of having a $U(1)$-symmetry even in presence of non-rotating wave (anti-resonant) coupling terms.  We determine the rich phase diagram of such model and show the appearance of Goldstone and amplitude (Higgs-like) mode on a line of the phase diagram. We show one example of circuit QED configuration where this kind of quantum model can be implemented, by using capacitive and inductive coupling of a Josephson junction artificial atom to a superconducting resonator.

{\it The Model -} The model we introduce here describes a collection of $N$ two-level systems, each one interacting with both the two quadratures of a bosonic mode (e.g. the electric and the magnetic field of an electromagnetic field). Namely, we consider the Hamiltonian
\begin{eqnarray}
\mathcal{H}&=&\hbar\omega_0 J_z + \hbar \omega a^{\dagger}a + \frac{\hbar \Omega_E}{\sqrt{N}}(a+a^{\dagger})(J_++J_-)\nonumber \\ &&+  \frac{\hbar \Omega_M}{\sqrt{N}}(a-a^{\dagger})(J_+-J_-) \label{FullHamiltonian} ,
\end{eqnarray}
where $\Omega_E$ and $\Omega_M$ are the coupling constants, $\omega$ represents the frequency of the bosonic mode, while $\omega_0$ is the transition frequency of each two-level system. The angular momentum operator represents the collective pseudo-spin associated to the collection of $N$ two-level systems ($J_+=\frac{1}{2}\sum_i\sigma_+^i$, $J_-=\frac{1}{2}\sum_i \sigma_-^i$, $J_z=\frac{1}{2}\sum_i\sigma_z^i$).
Note that the Hamiltonian terms proportional to $a^{\dagger} J_+$ and $a J_-$, which do not conserve the number of bare excitations, are the so-called non-rotating wave  coupling terms. These terms in general are responsible for denying a {\it continuous} symmetry to the Hamiltonian. However, since they describe the simultaneous creation or destruction of {\it two} excitations, the {\it parity} of the excitation number is conserved.  \\

\begin{figure}[t!]
   \begin{center}
   \includegraphics[width=180pt]{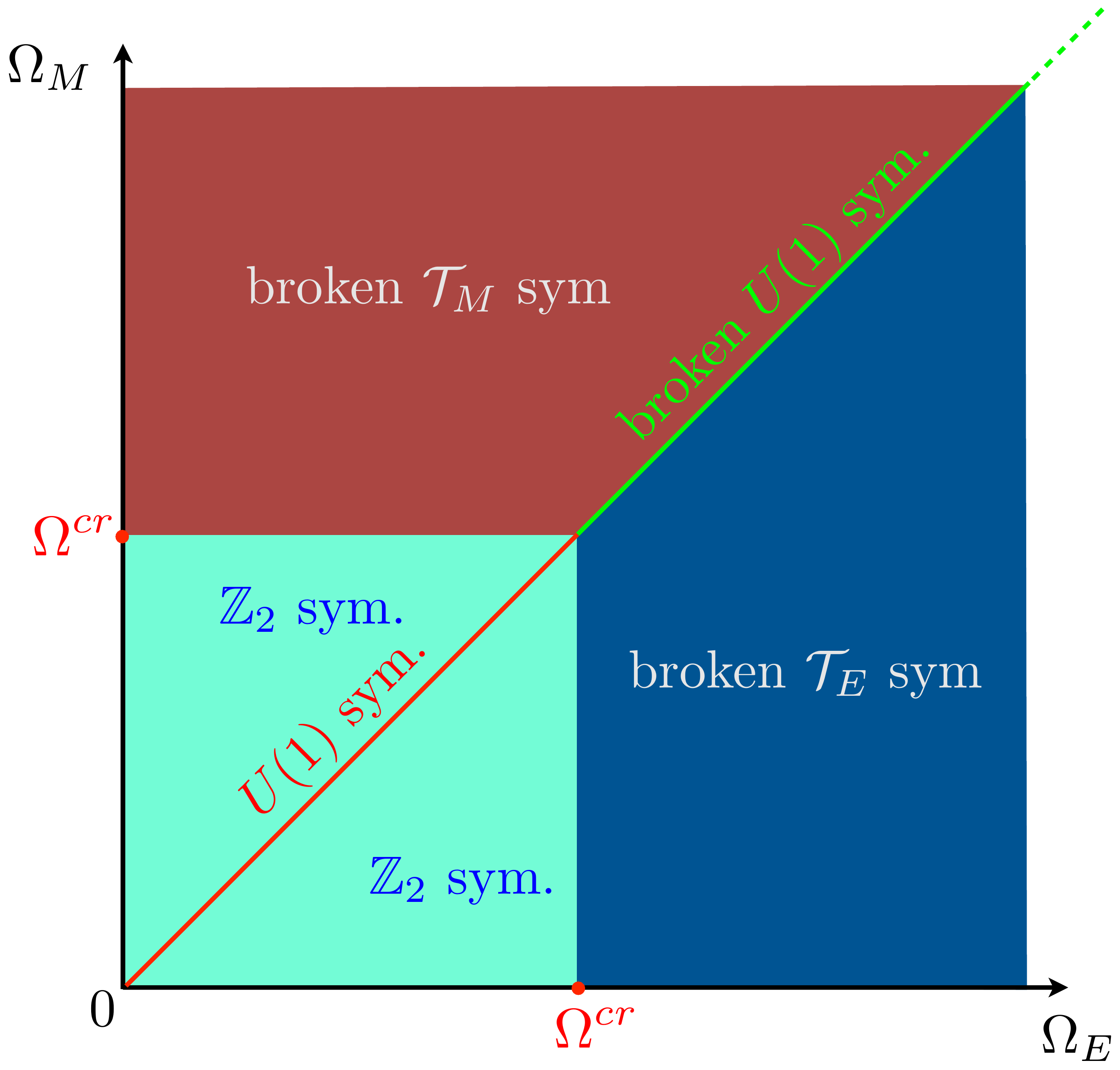}
   \end{center}
   \caption{(Color online) Symmetry diagram of the model described by Eq. (1) in the ($\Omega_E$,$\Omega_M$) plane \label{SymDiag}. The broken symmetries are due to superradiant phase transitions.
   The definition of the critical coupling constants are in the text.}
\end{figure}

{\it Symmetries -}  
When $\Omega_M\neq\Omega_E$, the Hamiltonian in Eq. (1) possesses a discrete $\mathbb{Z}_2$, parity symmetry \cite{Brandes}, which is composed of two other symmetries $\Pi=\mathcal{T}_E \circ \mathcal{T}_M$ that can be broken separately : 
\begin{align}
&\left(a+a^{\dagger},i(a-a^{\dagger}),J_x,J_y\right)\xrightarrow{\mathcal{T}_E} \left(-a-a^{\dagger},i(a-a^{\dagger}),-J_x,J_y\right)\nonumber\\
&\left(a+a^{\dagger},i(a-a^{\dagger}),J_x,J_y\right)\xrightarrow{\mathcal{T}_M}  \left(a+a^{\dagger},-i(a-a^{\dagger}),J_x,-J_y\right).
\end{align}
However, when we tune the couplings in such a way that $\Omega_M=\Omega_E=\Omega$ the Hamiltonian becomes :
\begin{eqnarray}
\mathcal{H}=\hbar\omega_0 J_z + \hbar \omega a^{\dagger}a + 2\frac{\hbar \Omega}{\sqrt{N}}(aJ_++a^{\dagger}J_-)\label{TCHamiltonian} .
\end{eqnarray}
Importantly,  non-rotating wave Hamiltonian terms cancel out in this Hamiltonian. Indeed, Eq. (\ref{TCHamiltonian}) is an Hamiltonian of the Tavis-Cummings \cite{TavisCummings} type which possesses a $U(1)$ symmetry characterized by the action of the operator $\mathcal{R}_{\theta}=\exp{(i\theta(a^{\dagger}a+J_z))}$:
\begin{align}
\mathcal{R}^{\dagger}_{\theta}\left(a,a^{\dagger},J_+,J_-\right)\mathcal{R}_{\theta}=(ae^{-i\theta},a^{\dagger}e^{i\theta},J_+ e^{i\theta},J_- e^{-i\theta}).
\end{align}
Hence, depending on the coupling ($\Omega_E$ and $\Omega_M$) it is possible to tune the symmetry of the model and have a continuous symmetry even if we have consistently considered the
non-rotating wave coupling terms. The symmetries of the considered Hamiltonian are summarized in Fig. \ref{SymDiag}. The occurrence of symmetry breaking is due to phase transitions, which are characterized in the following.

{\it Phase diagram -} In order to calculate the phase diagram of the model in Eq. (1),  we have used the Holstein-Primakoff approach, by considering the transformation ($J_+  =  b^{\dagger}\sqrt{N-b^{\dagger}b}\,\,,\,\,J_-=\sqrt{N-b^{\dagger}b}\,b\,\,,\,\,J_z = b^{\dagger}b-\frac{N}{2}$) that represents angular momentum operators in terms of bosonic operators $b$  and $b^{\dagger}$  in the Hamiltonian (\ref{FullHamiltonian}). Then, we have followed a mean-field approach by shifting the bosonic operators with respect to their mean value \cite{Brandes} ($a\rightarrow\alpha + c$, $b\rightarrow\beta + d$ with $\alpha=\langle a \rangle\propto\sqrt{N}$ and $\beta=\langle b \rangle\propto\sqrt{N}$). By keeping only the terms proportional to $N$ we obtain the mean value of the ground state energy in terms of $\alpha$, $\alpha^*$, $\beta$ and $\beta^*$ :
\begin{align}
E_G/\hbar =&\omega\left|\alpha\right|^2+\omega_0\left|\beta\right|^2+\Big{[}\Omega_E(\alpha+\alpha^*)(\beta+\beta^*)\nonumber\\
&+\Omega_M(\alpha-\alpha^*)(\beta^*-\beta)\Big{]}\sqrt{1-\frac{\left|\beta\right|^2}{N}}.\label{EG}
\end{align}
If there is a non-zero value of $\alpha$ and $\beta$ minimizing the energy, it means that the ground state of the system has a non-zero coherence of the boson field and a spontaneous pseudo-spin polarization of the two-level systems. Those coherences are the order parameters of the 'superradiant' phase transition for this model. The minimization of $E_G$ with respect to $\alpha^*$ leads to: 
\begin{align}
\alpha=-\left(\frac{\Omega_E}{\omega}(\beta+\beta^*)+\frac{\Omega_M}{\omega}(\beta-\beta^*)\right)\sqrt{1-\frac{\left|\beta\right|^2}{N}}\label{alpha}.
\end{align}
By substituting  in Eq. (\ref{EG}), we obtain the expression of the ground state energy in terms of $\beta$ and $\beta^*$ only ($E_G(\beta,\beta^*)$). A subsequent minimization of this function with respect to $\beta$ and $\beta^*$ completes the job to the determine the ground state coherence. 

\begin{figure}[t!]
   \begin{center}
   \includegraphics[width=180pt]{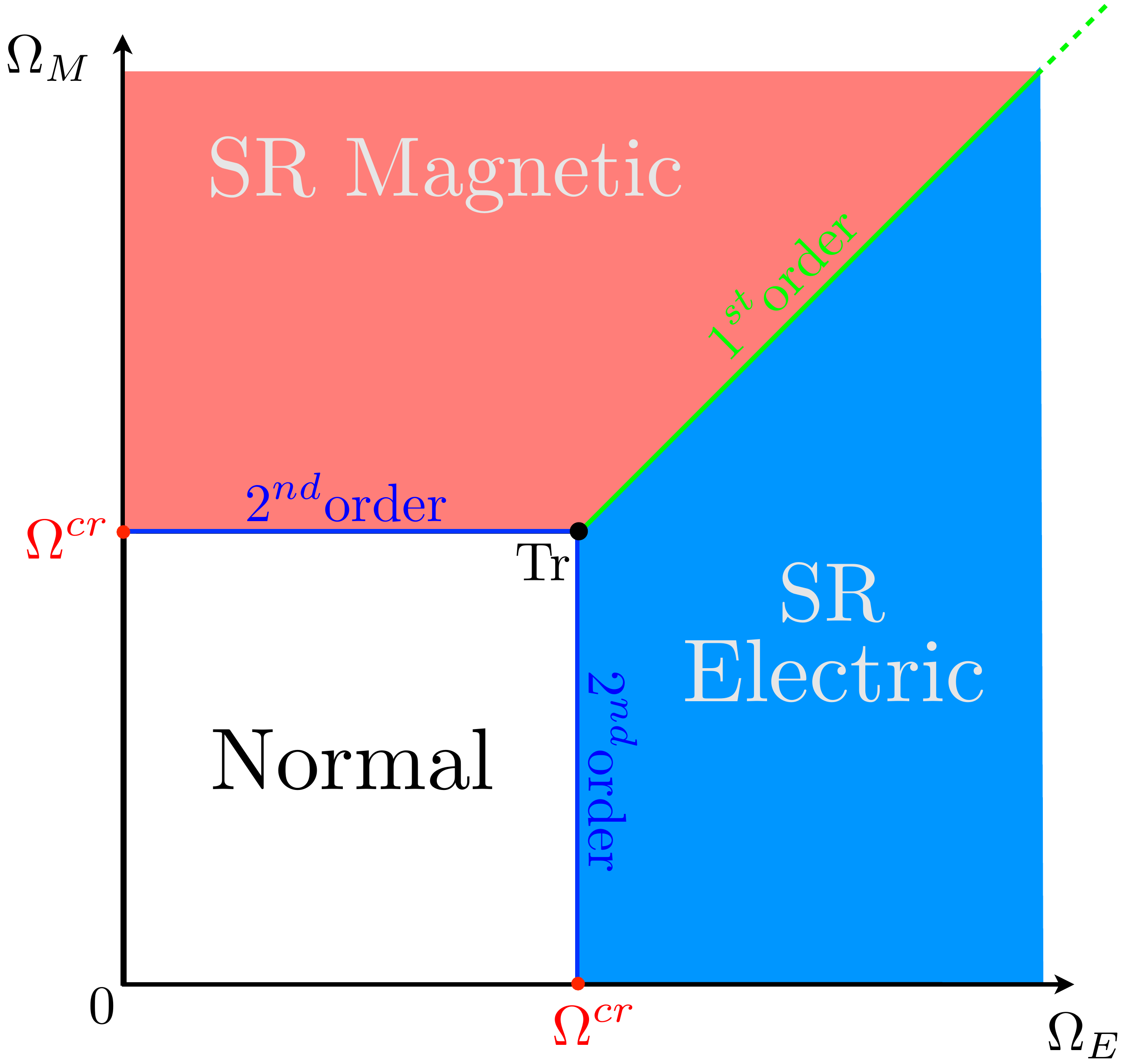}
   \end{center}
   \caption{(Color online) Phase diagram of the model in the ($\Omega_E$,$\Omega_M$) plane (thermodynamic limit).\label{PhaseDiag}}
\end{figure}

\begin{figure}[t!]
   \begin{center}
   \includegraphics[width=250pt]{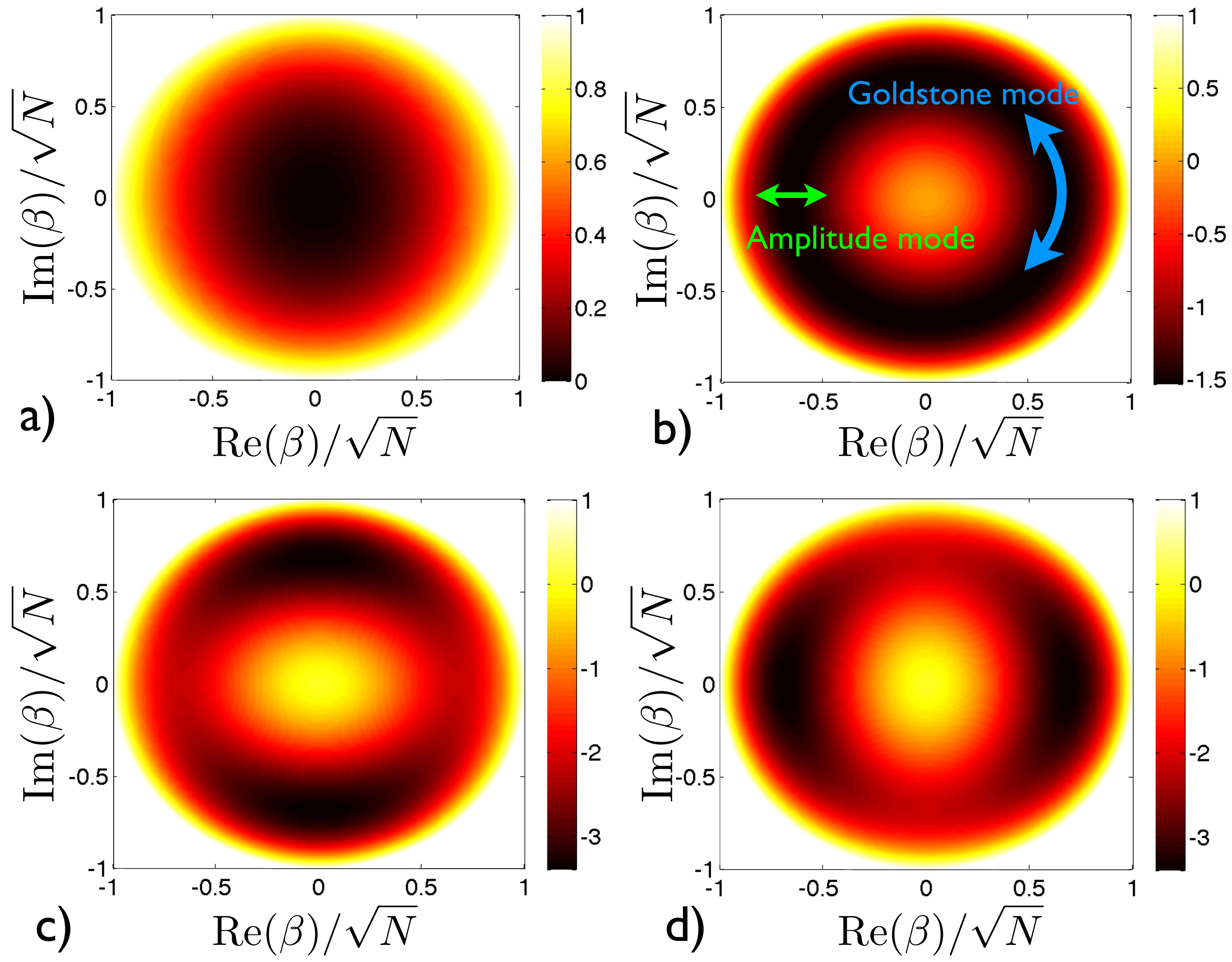}
   \end{center}
   \caption{(Color online) Ground state energy $E_G(\beta,\beta^*)/N$ in the ($\operatorname{Re}(\beta)/\sqrt{N}$,$\operatorname{Im}(\beta)/\sqrt{N}$) plane at resonance ($\omega=\omega_0$, $\Omega^{cr}=0.5  \omega$) for four different cases. {\bf a)} $\Omega_E=0.1 \omega$ and $\Omega_M=0.3 \omega$: the system is in the normal phase, with a minimum in the origin. {\bf b)} $\Omega_E=\Omega_M= \omega$: the ground state energy has  the shape of a Mexican hat, with a circular valley of degenerate minima. {\bf c)} $\Omega_E=1\omega$ and $\Omega_M=1.5\omega$: the energy is anistropic, with two minima in the imaginary axis. {\bf d)} $\Omega_E=1.5 \omega$ and $\Omega_M=1\omega$: two minima in the real axis. \label{GroundState}}
\end{figure}

To characterize the phases of the system, it is convenient to introduce the following  quantities: $\mu_E=\omega\omega_0/4\Omega_E^2$ , $\mu_M=\omega\omega_0/4\Omega_M^2$ and $\Omega^{cr}=\sqrt{\omega\omega_0}/2$. Our solutions show that there are four different regions in the phase diagram versus the coupling constants :\\

i) A 'normal' phase (see Fig. \ref{GroundState}a) is obtained for $\Omega_E<\Omega^{cr}$ and $\Omega_M<\Omega^{cr}$. In the normal phase, there is no ground state bosonic coherence and no pseudospin polarization $\left((\alpha,\beta)=(0,0)\right)$.\\

ii) A phase, which we call superradiant 'Electric' phase (see Fig. \ref{GroundState}d) is obtained for $\Omega_E>\Omega^{cr}$ and $\Omega_E>\Omega_M$. The ground state possesses a real bosonic coherence. The expressions for the order parameters are  $\left((\alpha, \beta)=(\mp \frac{\Omega_E}{\omega_0}\sqrt{N(1-\mu_E^2)}, \pm \sqrt{\frac
{N}{2}(1-\mu_E)})\right)$. Note that $\beta$ real means a pseudospin polarization along the $x$-direction ($\langle J_x \rangle \neq 0$). This phase breaks the $\mathcal{T}_E$ symmetry.\\

iii) A superradiant 'Magnetic' phase (see Fig. \ref{GroundState}c) is achieved for $\Omega_M>\Omega^{cr}$ and $\Omega_M>\Omega_E$, where the ground state possesses an imaginary bosonic coherence. The order parameters are $\left((\alpha, \beta)=(\mp i\frac{\Omega_M}{\omega_0}\sqrt{N(1-\mu_E^2)}, \pm i\sqrt{\frac{N}{2}(1-\mu_E)})\right)$. The pseudospin polarization is along the $y$-direction ($\langle J_y \rangle \neq 0$). Such a phase breaks the $\mathcal{T}_M$ symmetry.\\

iv) A superradiant 'EM' phase (fig. \ref{GroundState}b) is obtained for $\Omega_M=\Omega_E$ and $\Omega_E>\Omega^{cr}$. Here the ground state possesses a  complex bosonic coherence, while the pseudospin polarization is along the $\theta$ direction. The order parameters are $\left((\alpha, \beta)=(-\frac{\Omega_E}{\omega_0}\sqrt{N(1-\mu_E^2)}e^{i\theta},\sqrt{\frac{N}{2}(1-\mu_E)})e^{i\theta}\right)$. Such superradiant phase breaks the $U(1)$ symmetry.\\
 The behavior of the ground state energy as a function of the order parameters
is reported for four representative cases in Fig. \ref{GroundState}, in particular showing what happens in the normal phase (panel a), in the EM phase (panel b), in the 'Electric' phase (panel c) and in the 'Magnetic' phase (panel d).
In Fig. \ref{GroundState}(b) the ground state energy has a Mexican hat profile which is the visual manifestation of the broken $U(1)$ symmetry.
\\
\begin{figure}[t!]
   \begin{center}
   \includegraphics[width=260pt]{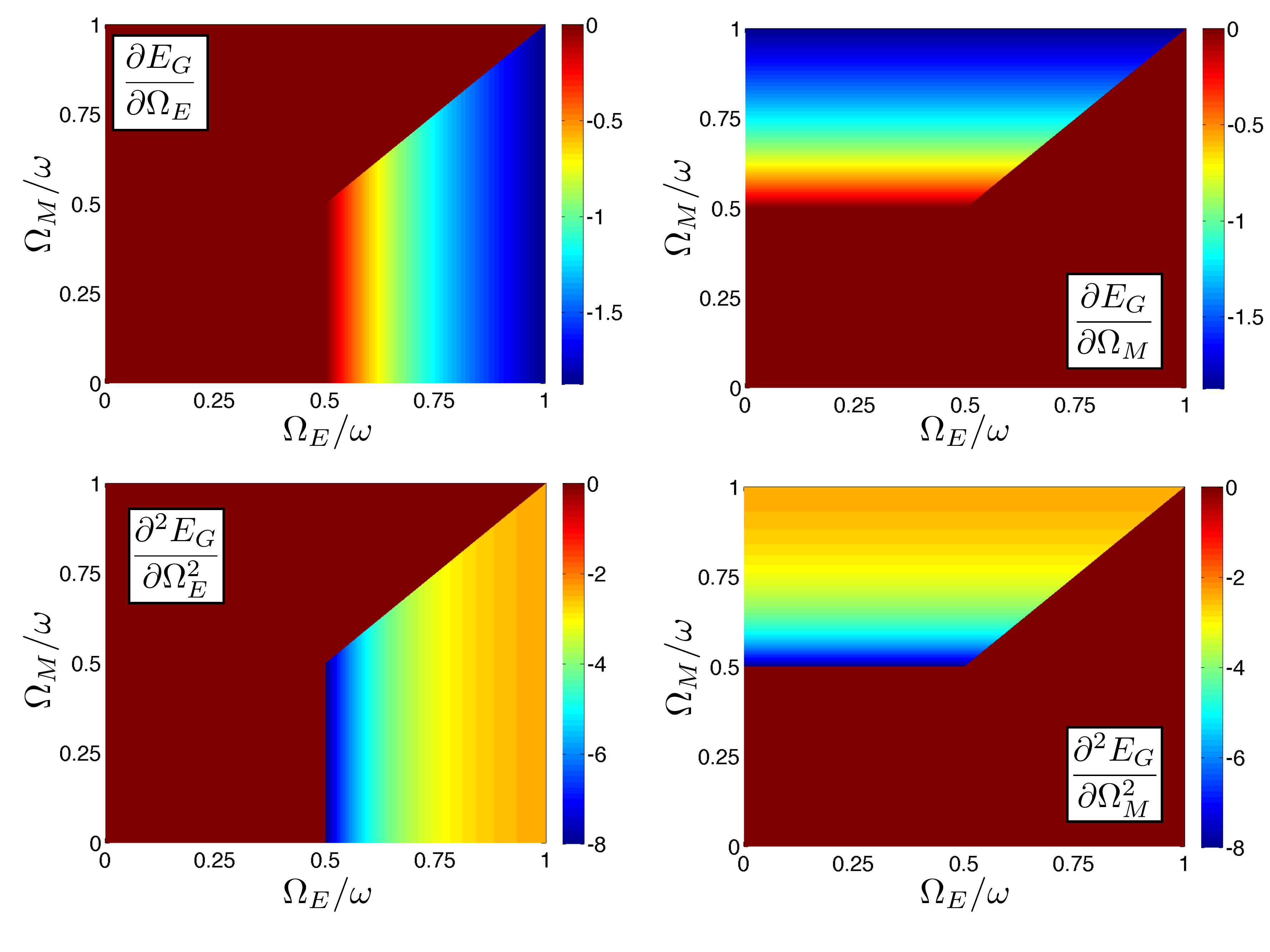}
   \end{center}
   \caption{(Color online) First (top panels) and second (bottom panels) derivative of the ground state energy with respect to $\Omega_E$ (left panels) and $\Omega_M$(right panels) in the ($\Omega_E,\Omega_M$) plane.
   The top panels show a diagonal line of first-order transition points. The bottom panels indicate horizontal and vertical lines of second-order transition points.}\label{Derivative} 
\end{figure}

In order to determine the type of phase transitions, we have studied the discontinuities of the ground state energy at the transition points.
Since $\partial^2 E_G/\partial\Omega_E^2$ is discontinuous at $\Omega_E=\Omega^{cr}$ and $\Omega_E>\Omega_M$ (see Fig. \ref{Derivative}) the transition from the normal state to the superradiant Electric state is of second order. Analogously, the transition from the normal to the superradiant Magnetic phase is also of second order. On the other hand, since $\partial E_G/\partial\Omega_E$  and $\partial E_G/\partial\Omega_M$ are both discontinuous at $\Omega_E,\Omega_M>\Omega^{cr}$ and $\Omega_E=\Omega_M$, the transition from superradiant Electric to superradiant Magnetic is of first order.\\

{\it Excitation spectra -} By generalizing the approach in Ref. \cite{Brandes}, it is possible to obtain the energies of the bosonic excitations in each phase. In particular, there are two 'polariton' bosonic excitation branches:
$\epsilon_+$ ($\epsilon_-$) stands for the upper (lower) polariton branch energy. The analytical expressions read:
\begin{align}
(\epsilon_{\pm} / \hbar)^2= & \frac{1}{2}\Big{\{}8\tilde{\Omega}_E\tilde{\Omega}_M + \omega^2+\tilde{\omega}_0^2 \pm \sqrt{(\omega^2-\tilde{\omega}_0^2)^2} \nonumber \\  & \overline{ +16(\tilde{\Omega}_E \tilde{\omega}_0+\tilde{\Omega}_M\omega)(\tilde{\Omega}_E \omega+\tilde{\Omega}_M \tilde{\omega}_0)}\Big{\}},
\end{align}
with $(\tilde{\omega}_0,\tilde{\Omega}_E,\tilde{\Omega}_M)=(\omega_0,\Omega_E,\Omega_M)$ in the Normal phase, $(\tilde{\omega}_0,\tilde{\Omega}_E,\tilde{\Omega}_M)=(\omega_0/\mu_E,\Omega_E\mu_E,\Omega_M)$ in the superradiant Electric phase and $(\tilde{\omega}_0,\tilde{\Omega}_E,\tilde{\Omega}_M)=(\omega_0/\mu_M,\Omega_E,\Omega_M\mu_M)$ in the  superradiant Magnetic phase.

\begin{figure}[t!]
   \begin{center}
   \includegraphics[width=180pt]{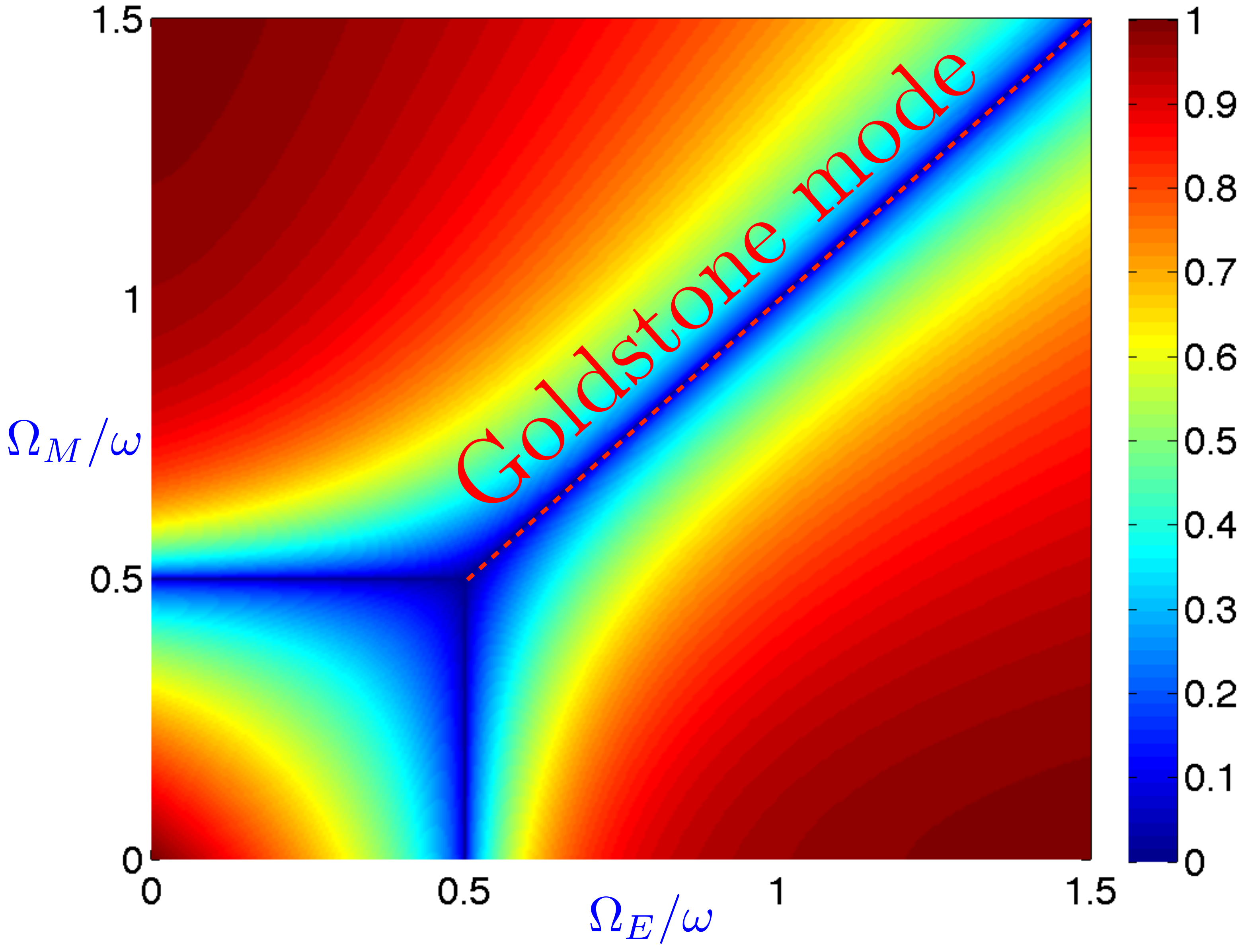}  
    \includegraphics[width=180pt]{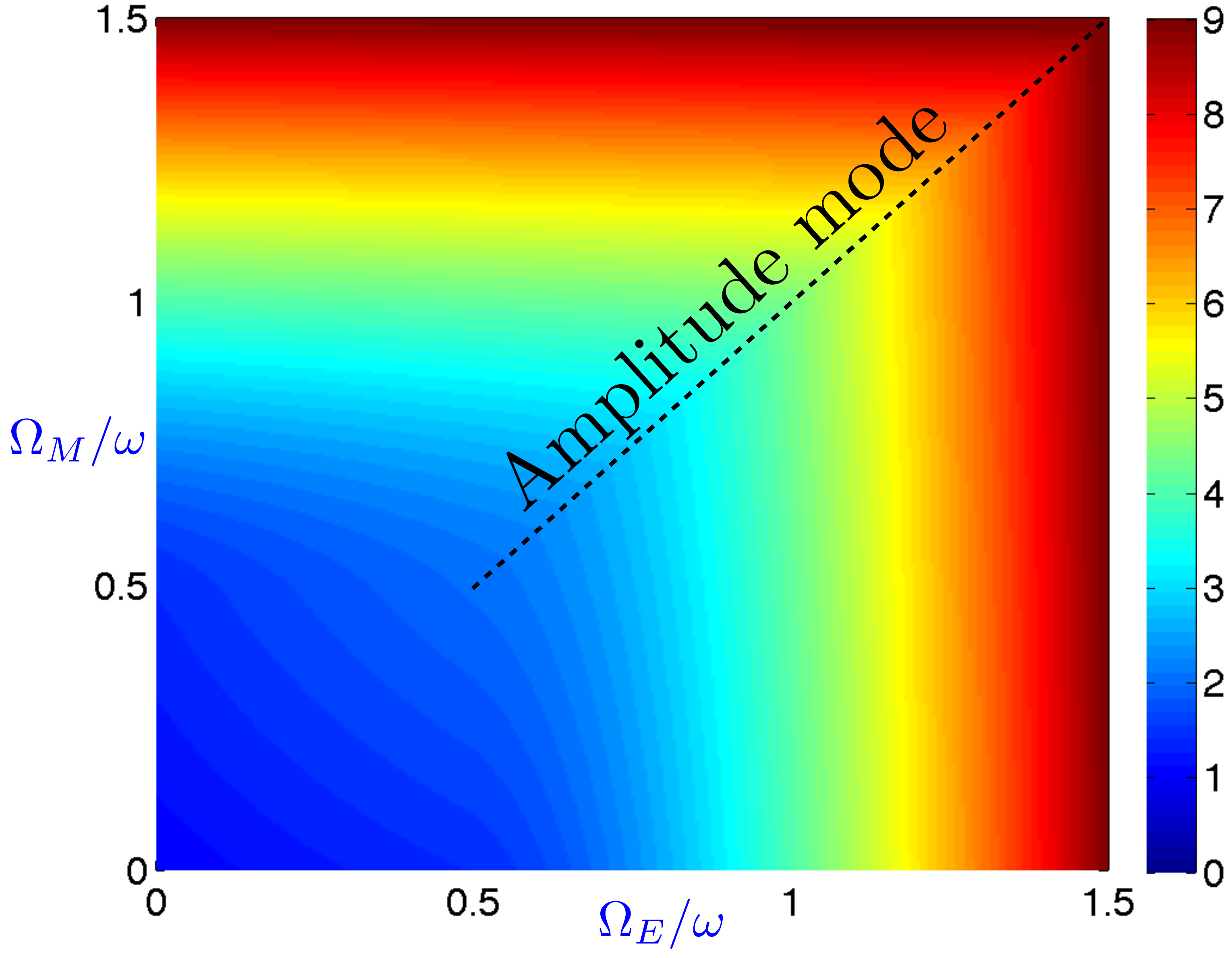}
   \end{center}
   \caption{(Color online) Lower ($\epsilon_-$, bottom panel) and upper ($\epsilon_+$, top panel) polariton branches energy (in units of $\hbar \omega$) on resonance ($\omega=\omega_0$) in the ($\Omega_E$,$\Omega_M$) plane (thermodynamic limit).\label{Emoins}}
\end{figure}

The results for the energy of the lower polariton branch have been plotted in the top panel of Fig. (\ref{Emoins}). Note that there are three critical lines where the lower polariton branch energy is zero (gapless excitation). We also underline the presence of a Goldstone mode when $\Omega_E=\Omega_M > \Omega^{cr}_{M}$ due to the breaking of the continuous $U(1)$ symmetry. The behavior of the upper polariton branch is reported in the bottom panel of  Fig. (\ref{Emoins}), showing a  finite-energy amplitude (Anderson-Higgs-like) mode in the diagonal line of the phase diagram where the Goldstone mode occurs. \\

{\it Implementation in circuit QED systems -} The rich model explored in this letter can be implemented by using circuit QED systems (see Fig. \ref{circuit}). The superconducting circuit consists of a collection of Josephson junction artificial atoms coupled both inductively and capacitively to a transmission line resonator. In the case of circuit QED, the phase operator $\varphi$ and the number operator $N$ are conjugate ($\left[\varphi,N\right]=i$),  playing a role similar to position and momentum for mechanical degrees of freedom.  The Hamiltonian describing the circuit in Fig. \ref{circuit} reads:
\begin{figure}[t!]
   \begin{center}
   \includegraphics[width=250pt]{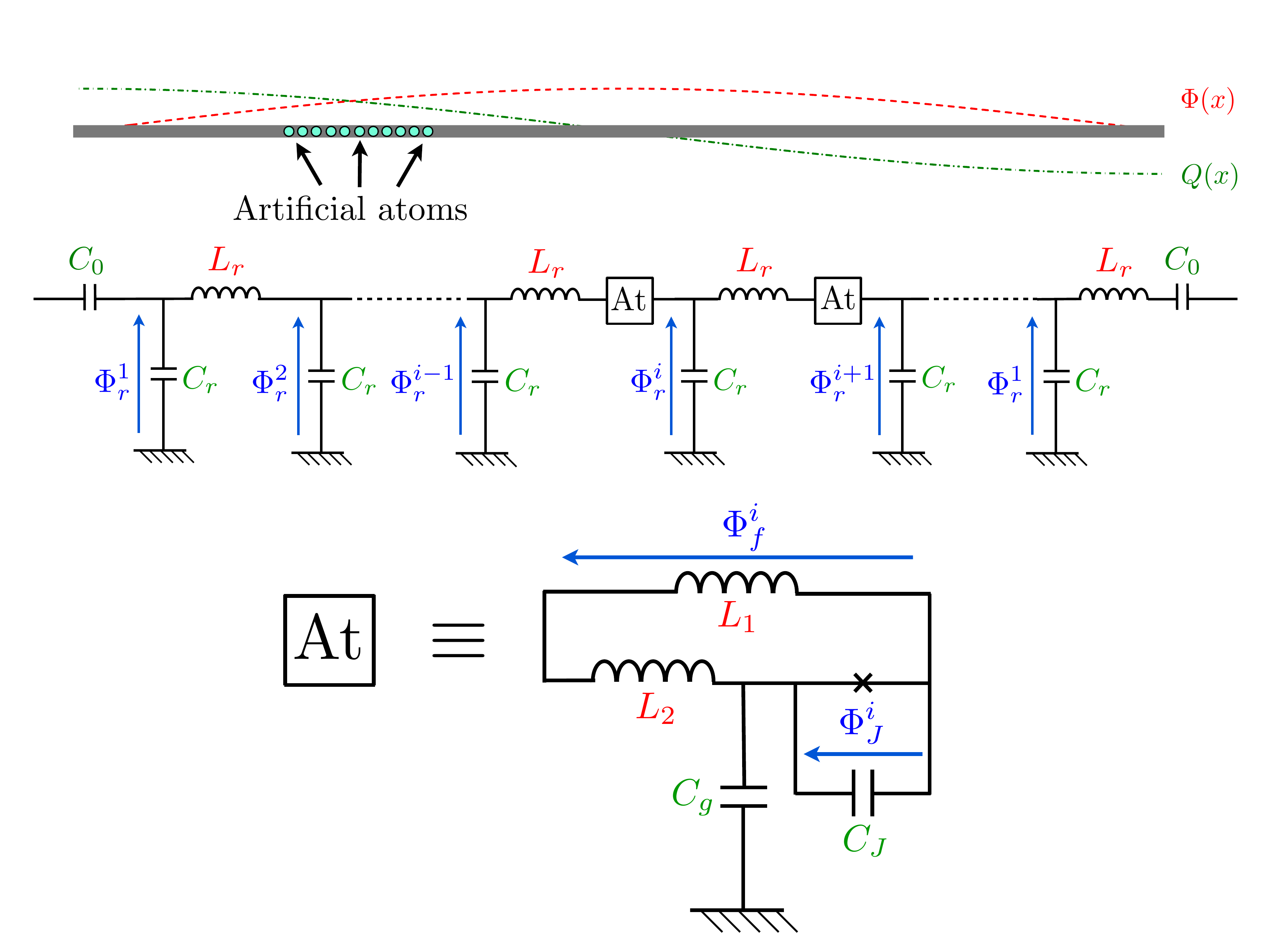}
   \end{center}
   \caption{\label{circuit}(Color online) Design of circuit QED system for the implementation of the Hamiltonian (\ref{FullHamiltonian}). Each artificial atom
(obtained with a Josephson junction) is coupled both capacitively and inductively to a transmission line resonator (equivalent to a series of LC resonators). }
\end{figure}

\begin{align}
\mathcal{H}_{circ}=&\sum_i\Big{\{}E_{C_r}(N_r^i)^2+E_{L_r}(\varphi_r^{i+1}-\varphi_r^{i})^2\nonumber\\
&+E_{C_J}(N_J^i)^2+E_{L_J}(\varphi_J^i)^2-E_J\cos{(\varphi_J^i+\varphi_{ext})}\nonumber\\
&+G_Q \,\,N_J^i N_r^{i+1}+G_L\,\,\varphi_J^i(\varphi_r^{i+1}-\varphi_r^i)\Big{\}}.
\end{align}
For $C_r>>C_g,C_J$ and $L_r>>L_1,L_2$, we have $E_{C_r}=2e^2/C_r$, $E_{L_r}=(h/2e)^2/2L_r$, $E_{C_J}=2e^2/(C_g+C_J)$, $E_{L_J}=(h/2e)^2(L_r+L_1)/2(L_1+L_2)$, $G_Q=4e^2C_g/C_r(C_g+C_J)$, $G_L=(h/2e)^2L_1/L_r(L_1+L_2)$.
By quantizing the resonator modes \cite{Blais2004}, keeping only one resonator mode and doing a two-level system approximation for the artificial atoms (quasi-resonant to the resonator mode) at the sweet spot of the Josephson atomic Hamiltonian ($\varphi_{ext}=\pi$), we obtain :
\begin{align}
\mathcal{H}_{circ}/\hbar =&\omega_{res}a^{\dagger}a+\omega_{J}J_z+\sum_j\tilde{G}_E^j(a+a^{\dagger})\sigma_x^j\nonumber\\
&+i\sum_j\tilde{G}_M^j(a-a^{\dagger})\sigma_y^j.
\end{align}
If the artificial atoms are identically coupled ($\tilde{G}_E^j = \tilde{G}_E$) and ($\tilde{G}_M^j = \tilde{G}_M$),  we recover the Hamiltonian (\ref{FullHamiltonian}).
Of course, this is not the only possible implementation. Importantly, it shows that ultrastrong coupling circuit QED can give access to this kind of physics.

In conclusion, we have studied a new paradigm of model where two-level artificial atom systems are coupled to both quadratures of a bosonic field.
In such a model, which includes non-rotating wave terms of the atom-field coupling, it is possible to control the symmetries of the Hamiltonian in a remarkable way,
with the possibility of having a $U(1)$ continuous symmetry even in the ultrastrong coupling regime. The phase diagram presents 4 types of superradiant phases, with one phase having Goldston gapless excitations
and amplitude mode excitations. We have shown that by using circuit QED systems it is possible to implement this double quadrature coupling scheme, paving the way to the exploration of  rich spontaneous symmetry breaking physics in photonic systems. Our theoretical paradigm could also stimulate implementations with driven superfluid Bose-Einstein condensates in optical cavities\cite{RitschRMP}.

We thank Pierre Nataf and Enrique Solano for discussions. We acknowledge support from ANR grant QPOL. C. C. is member of Institut Universitaire de France.

\end{document}